# Graphical Abstract

## A new LDA formulation with covariates.

Gilson Y. Shimizu,Rafael Izbicki,Denis Valle

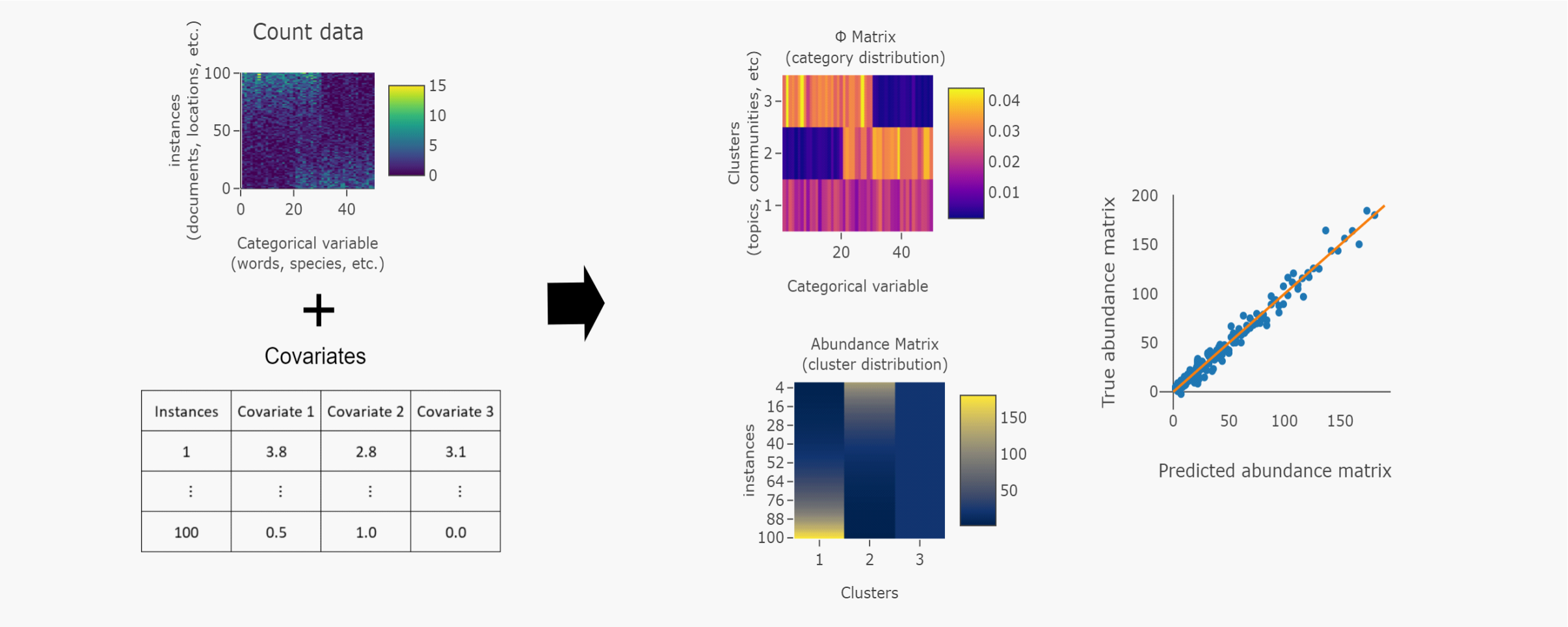

# Highlights

## A new LDA formulation with covariates.

Gilson Y. Shimizu,Rafael Izbicki,Denis Valle

- We propose a new formulation for the Latent Dirichlet Allocation (LDA) model which incorporates covariates.
- Our extension enables a straight-forward interpretation of the regression coefficients and the analysis of the quantity of cluster-specific elements in each sampling unit - including the prediction of these quantities in a new sample through covariates.
- We illustrate the benefits of this formulation using three data sets: text-mining of Coronavirus articles, analysis of grocery shopping baskets, and ecology of tree species in Barro Colorado Island.
- We provide an R package that enables users to readily apply our model.

# A new LDA formulation with covariates.

Gilson Y. Shimizu[a,*], Rafael Izbicki[b,*] and Denis Valle[c,*]

[a]Institute of Mathematics and Computer Sciences, University of Sao Paulo, Sao Paulo, Brazil.
[b]Department of Statistics, University Federal de Sao Carlos, Sao Paulo, Brazil.
[c]School of Forest Resources and Conservation, University of Florida, Gainesville, Florida, United States of America.

ABSTRACT

The Latent Dirichlet Allocation (LDA) model is a popular method for creating mixed-membership clusters. Despite having been originally developed for text analysis, LDA has been used for a wide range of other applications. We propose a new formulation for the LDA model which incorporates covariates. In this model, a negative binomial regression is embedded within LDA, enabling straight-forward interpretation of the regression coefficients and the analysis of the quantity of cluster-specific elements in each sampling units (instead of the analysis being focused on modeling the proportion of each cluster, as in Structural Topic Models). We use slice sampling within a Gibbs sampling algorithm to estimate model parameters. We rely on simulations to show how our algorithm is able to successfully retrieve the true parameter values and the ability to make predictions for the abundance matrix using the information given by the covariates. The model is illustrated using real data sets from three different areas: text-mining of Coronavirus articles, analysis of grocery shopping baskets, and ecology of tree species on Barro Colorado Island (Panama). This model allows the identification of mixed-membership clusters in discrete data and provides inference on the relationship between covariates and the abundance of these clusters.

## 1. Introduction

Unsupervised machine learning methods allow the analysis of multivariate data sets in which no response variable is available. This type of analysis is especially useful as the amount of unstructured information grows (in the form of texts, for example), enabling the unveiling of latent structure in the data. In particular, the Latent Dirichlet Allocation (LDA) method is an unsupervised technique that focuses on identifying unobservable groups. This method is different from traditional unsupervised methods such as hard clustering, where sampling units can only be classified into a single group. In LDA, soft clustering is performed, that is, a sample unit can belong to several groups at the same time.

The LDA model was originally proposed by Pritchard et al. [18] in the context of population genetics, but it became popular in the context of machine learning through the work of Blei et al. [3] on text-mining applications. In these text-mining applications, the goal is to discover topics that are present in each document based on the words that appear on these documents. This model has been applied to several areas of knowledge. For example, Lukins et al. [14] used LDA to understand software bug reports. In another application, Lienou et al. [12] applied this model to create annotation of satellite imagery. LDA was also used by Xing and Girolami [22] to detect fraudulent calls in the telecommunications industry based on the patterns found for each customer. Finally, Valle et al. [20] used LDA on biodiversity data to describe groups of trees.

*Corresponding author
gilsonshimizu@yahoo.com.br (G.Y. Shimizu); rafaelizbicki@gmail.com (R. Izbicki); drvalle@ufl.edu (D. Valle)

Several variations of LDA exist. For instance, Mcauliffe and Blei [15] introduced the supervised LDA model where documents are labeled with continuous or discrete response variables. Wang and Grimson [21] considered a spatial structure to group spatially close elements (such as words that are close in the text). Blei and Lafferty [2] analyzed the evolution of topics over time through a family of probabilistic time series models. Albuquerque et al. [1] adapted the LDA model for different types of data (multinomial, binomial and bernoulli) and used a special prior called truncated stick-breaking (TSB) prior to identify the optimal number of groups.

In many problems, one also has access to additional information about instances that comes in the form of features (covariates). For example, a company may have socioeconomic information about its customers, such as age or income, that can help in understanding customer interactions via chat. In these cases, it can be useful to explore the relationship between these covariates and the identified groups.

**Related work**. Roberts et al. [19] developed Structural Topic Models (STM), in which covariates were incorporated into LDA through a Multinomial regression model so that the probability of each topic in a given instance is allowed to depend on covariates. The focus on the probability of each topic, instead of the abundance of each topic, is an important limitation. For example, the type of scientific article (e.g., a commentary or a review article) can significantly change the number of words associated with each topic. However, STM's might fail to identify this effect if the proportion of the different topics remains the same. Like STM, Jacobs et al. [9] also considers covariates to explain the proportions of each cluster through a logarithmic link function but with a focus on predictions, not inference. Furthermore, the approach proposed in Jacobs et al. [9] requires running the model multiple times to determine the optimal number of clusters. On the other hand, our approach can readily determine the optimal number of clusters by running the model just once and can be used both for prediction and for inference on regression coefficients.

In this work, we propose a new formulation to the LDA model where we use covariates to explain the number of elements (e.g., number of words) in each group, rather than the proportion of each group. In a sense, our model is more general than STM's because the probabilities of each group can also be derived from it. Another important advantage of our approach is that the covariate coefficients (i.e., the slope parameters) can be interpreted more easily through the logarithmic link function of the Negative-Binomial regression rather than the logistic link function used within the Multinomial regression in STM's. The log link function (Fig. 1 (a)) allows a straightforward interpretation of the coefficients in the sense that a positive (negative) coefficient describes a positive (negative) relationship between the corresponding covariate and the abundance of each group. On the other hand, as illustrated in Fig. 1 (b), a positive coefficient in the multinomial logistic link function can imply a positive or negative relationship between the proportion of the group and the corresponding covariate.

In section 2 we present the proposed Bayesian model and the full conditional distributions required for our Gibbs sampler. In section 3 we describe the estimation method and the software used. Then, in section 4, we apply our model to simulated data sets to demonstrate its effectiveness in providing inferences on the parameters of interest. In section 5, we illustrate the versatility of our model by applying it to data sets from different fields. Section 6 compares our model to STMs using the probabilistic coherence metric. We conclude with a discussion of the advantages and disadvantages of the method and suggestions for future research.

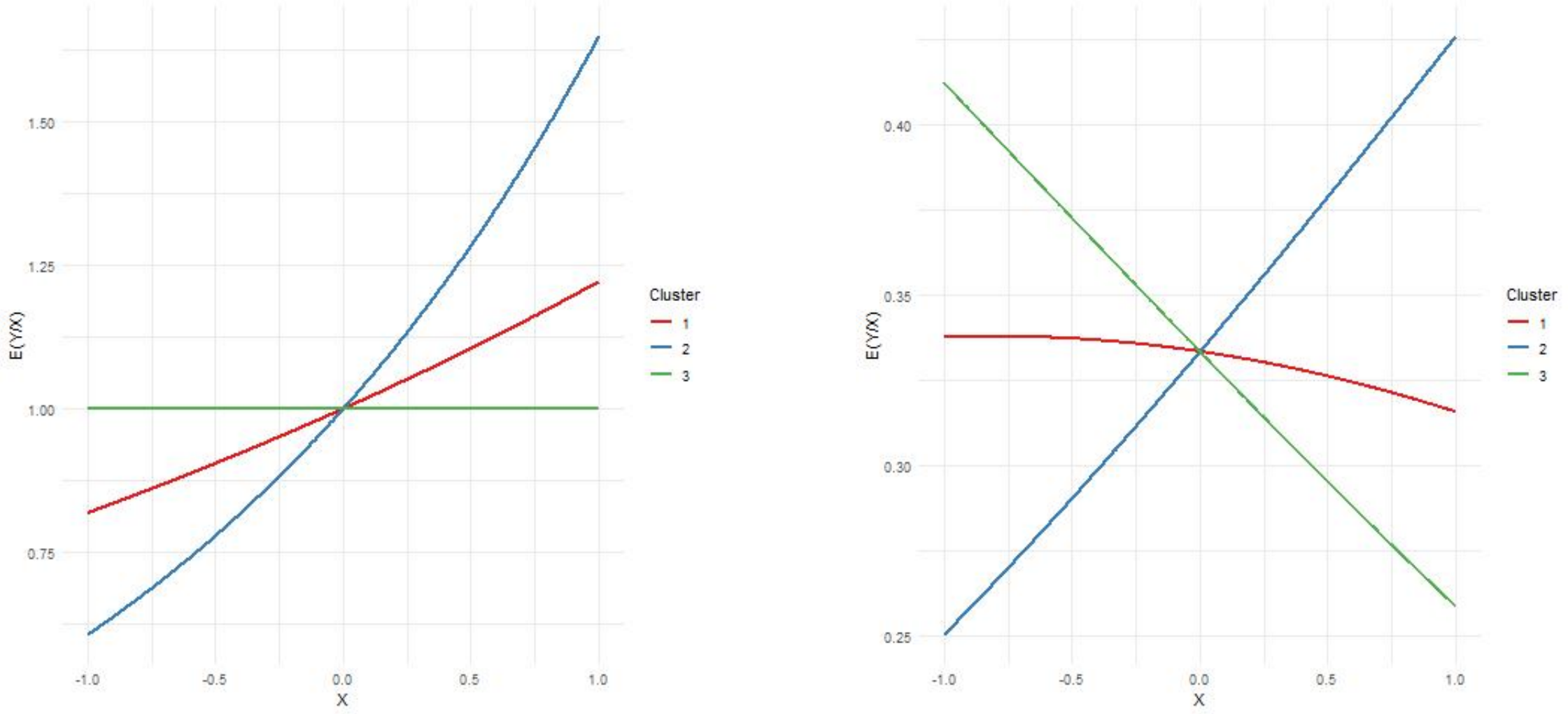

(a) Negative-binomial log link

(b) Multinomial logistic link

**Figure 1:** Illustration of the difference between the logarithmic (left panel) and multinomial logistic (right panel) link functions considering 3 groups and 1 covariate (without intercept) with $\beta_1 = 0.2$, $\beta_2 = 0.5$ and $\beta_3 = 0.0$. Notice that, despite the positive coefficient for group 1, there is a negative relationship in panel (b) between the covariate and the expected value of y given x. (a) Negative-binomial log link. (b) Multinomial logistic link.

## 2. Model

We start by introducing the proposed model. Let $L$ denote the total number of instances, $K$ be the total number of clusters and $S$ be the total number of categories each element from each instance can belong to. For example, in text analysis we can have $K$ topics, a vocabulary of $S$ distinct words and $L$ documents. In this case, the words in each document are the elements. We denote by $n_{l,*,*} := \sum_{s,k} n_{l,s,k}$ the total number of elements at instance $l$, where $n_{l,s,k}$ is a latent variable representing the total number of elements of category $s$ and cluster $k$ in instance $l$. The data that we observe consist of

- $y_{i,l} \in \{1, \dots, S\}$, the category of the $i$-th element of instance $l$, $i = 1, \dots, n_{l,*,*}$ and $l = 1, \dots, L$.
- $\mathbf{x}_l$: a $d$-dimension vector with the features (covariates) associated to instance $l$, $l = 1, \dots, L$.

The data $y_{i,l}$ are often summarized as an instance-by-category abundance matrix. More specifically, each cell in this matrix contains the total number of elements of category $s$ on instance $l$, given by $w_{l,s} := \sum_i \mathbb{1}(y_{i,l} = s)$, $l = 1, \dots, L$ and $s = 1, \dots, S$.

In our model, the link between the covariates and the abundance of each cluster in each instance $n_{l,*,k} = \sum_{s=1}^{S} n_{l,s,k}$ is given by a Negative-Binomial regression:

$$n_{l,*,k} \mid \boldsymbol{\beta}_k, N \sim \text{NegBinom}(\lambda_{l,k}, N)$$

where $\boldsymbol{\beta}_k$ is a $d$-dimension vector, $N$ is the overdispersion parameter and $\lambda_{l,k} := E[n_{l,*,k}] =: \exp(\mathbf{x}_l^T \boldsymbol{\beta}_k)$. Notice that $n_{l,*,k}$ are latent variables (and thus are not observed), and therefore we need to estimate the coefficients of the regression function *at the same time* we estimate $n_{l,*,k}$ (see details in Section 2.1).

The model also assumes that

$$(n_{l,1,k}, \dots, n_{l,S,k}) \mid n_{l,*,k}, \boldsymbol{\phi}_k \sim \text{Multinomial}(n_{l,*,k}, \boldsymbol{\phi}_k),$$

where $\boldsymbol{\phi}_k \in \mathbb{R}^S$ is a vector on a simplex that represents the composition of categories inside cluster $k$. We call $\Phi$ the matrix with elements $\phi_{k,s}$. Furthermore, note that within the standard LDA model, a parameter of primary interest is $\theta_{l,k}$, which is the proportion of cluster $k$ in instance $l$. This parameter can be easily retrieved based on the $n_{l,*,k}$ results from our model by calculating $\theta_{l,k} = \frac{n_{l,*,k}}{\sum_{c=1}^{K} n_{l,*,c}}$. We call $\Theta$ the matrix with elements $\theta_{l,k}$.

We use the following prior distributions:

$$N \sim \text{Unif}(0, N_0),$$

$$\boldsymbol{\phi}_k \mid \boldsymbol{\gamma} \sim \text{Dirichlet}(\boldsymbol{\gamma}), \boldsymbol{\gamma} = (\gamma_1, \dots, \gamma_S),$$

and

$$\boldsymbol{\beta}_k \sim N_d(\mathbf{0}, \mathbf{T}),$$

where $\mathbf{T}$ is a diagonal matrix. The hyper parameters $N_0$, $\boldsymbol{\gamma}$ and $\mathbf{T}$ are a priori set by the modeler.

The joint density function induced by the likelihood function and prior distributions is given by

$$\begin{aligned} &p(\{n_{l,s,k}\}, \{\boldsymbol{\phi}_k\}, \{\boldsymbol{\beta}_k\} | \{w_{l,s}\}, \{x_l\}) \propto \\ &\left[ \prod_{l=1}^{L} \prod_{k=1}^{K} \left[ \text{Multinomial}([n_{l,1,k}, \dots, n_{l,S,k}] \mid n_{l,*,k}, \boldsymbol{\phi}_k) \text{NegBinom}(n_{l,*,k} \mid \exp(\mathbf{x}_l^T \boldsymbol{\beta}_k), N) \right]^{\mathbb{1}(w_{l,s} = \sum_{k=1}^{K} n_{l,s,k})} \right] \times \\ &\times \left[ \prod_{k=1}^{K} \text{Dirichlet}(\boldsymbol{\phi}_k \mid \boldsymbol{\gamma}) \right] \left[ \prod_{k=1}^{K} N_d(\boldsymbol{\beta}_k \mid \mathbf{0}, \mathbf{T}) \right] \text{Unif}(N \mid 0, N_0). \end{aligned}$$

## 2.1. Full Conditional Distributions

We can obtain samples from the posterior distribution by using a Gibbs sampler [7]. In order to do that, we first derive the full conditional distributions for the parameters in our model. First, we derive the conditional distribution of each $\boldsymbol{\phi}_k$ given all the other quantities:

$$
\begin{aligned}
p(\boldsymbol{\phi}_k \mid \dots) &\propto \left[\prod_{i=1}^{n_{l,*,*}} \prod_{l=1}^{L} \text{Categorical}(y_{il} \mid \boldsymbol{\phi}_k)^{\mathbb{1}(z_{il}=k)}\right] \text{Dirichlet}(\boldsymbol{\phi}_k \mid \boldsymbol{\gamma}) \\
&\propto \left[\prod_{i=1}^{n_{l,*,*}} \prod_{l=1}^{L} \phi_{k,1}^{\mathbb{1}(y_{il}=1, z_{il}=k)} \times \dots \times \phi_{k,S}^{\mathbb{1}(y_{il}=S, z_{il}=k)}\right] \phi_{k,1}^{\gamma_1 - 1} \times \dots \times \phi_{k,S}^{\gamma_S - 1} \\
&\propto \phi_{k,1}^{n_{*,1,k}+\gamma_1-1} \times \dots \times \phi_{k,S}^{n_{*,S,k}+\gamma_S-1}.
\end{aligned}
$$

Thus,

$$
\boldsymbol{\phi}_k \mid \dots \sim \text{Dirichlet}\left(\left[n_{*,1,k} + \gamma_1, \dots, n_{*,S,k} + \gamma_S\right]\right),
$$

which is straightforward to sample from.

The conditional distribution of $\boldsymbol{\beta}_k$ given all of the other quantities is

$$
p(\boldsymbol{\beta}_k \mid \dots) \propto \left[\prod_{l=1}^{L} \text{NegBinom}\left(n_{l,*,k} \mid \exp(\mathbf{x}_l^T \boldsymbol{\beta}_k), N\right)\right] N_d(\boldsymbol{\beta}_k \mid \mathbf{0}, \tau^2 I_d).
$$

Because of lack of conjugacy, we rely on a slice-sampler algorithm (see Appendix C) to sample from this FCD.

For the parameter $N$, we obtain

$$
p(N \mid \dots) \propto \left[\prod_{k=1}^{K} \prod_{l=1}^{L} \text{NegBinom}\left(n_{l,*,k} \mid \exp(\mathbf{x}_l^T \boldsymbol{\beta}_k), N\right)\right] Unif(N|0, N_0).
$$

Again, we rely on a slice-sampler algorithm to sample from this FCD.

Finally, we obtain the conditional distribution of $z_{i,l}$ (the latent group membership of the $i$-th element in the $l$-th instance):

$$
\begin{aligned}
&p(z_{i,l'} = k \mid y_{i,l'} = s', \dots) \\
&\propto \prod_{l=1}^{L} \prod_{k=1}^{K} \left[\text{Multinomial}([n_{l,1,k}, \dots, n_{l,S,k}] \mid n_{l,*,k}, \boldsymbol{\phi}_k) \text{NegBinom}(n_{l,*,k} \mid \lambda_{l,k}, N)\right]
\end{aligned}
$$

After integrating $\boldsymbol{\phi}_k$ out and simplifying this expression (a detailed derivation of these results is provided in Appendix

A and B), we obtain:

$$p(z_{i,l'} = k \mid y_{i,l'} = s', \dots) \propto \frac{(n_{l',*,k} + N)(n_{*,s',k} + \gamma_{s'})}{(n_{l',s',k} + 1)(n_{*,*,k} + \sum_s \gamma_s)}(1 - p_{l',k}).$$

where $p_{l',k} = \frac{N}{N+\lambda_{l',k}}$. Thus,

$$z_{i,l} \mid y_{i,l} = s, \cdots \sim \text{Categorical}\left(\left[\frac{\frac{(n_{l,*,1}+N)(n_{*,s,1}+\gamma_s)}{(n_{l,s,1}+1)(n_{*,*,1}+\sum_s \gamma_s)}(1 - p_{l,1})}{\sum_{k=1}^{K} \frac{(n_{l,*,k}+N)(n_{*,s,k}+\gamma_s)}{(n_{l,s,k}+1)(n_{*,*,k}+\sum_s \gamma_s)}(1 - p_{l,k})}, \dots, \frac{\frac{(n_{l,*,K}+N)(n_{*,s,K}+\gamma_s)}{(n_{l,s,K}+1)(n_{*,*,K}+\sum_s \gamma_s)}(1 - p_{l,K})}{\sum_{k=1}^{K} \frac{(n_{l,*,k}+N)(n_{*,s,k}+\gamma_s)}{(n_{l,s,k}+1)(n_{*,*,k}+\sum_s \gamma_s)}(1 - p_{l,k})}\right]\right),$$

which is easy to sample from.

## 3. Estimation and Software

In order to fit our model, we first need to define $K$, the number of clusters that will be used. We do this by using the LDA model proposed by [1]. Instead of relying on the standard but computationally inefficient approach of fitting multiple models (one for each K value) to then determine the optimal number of clusters, the model in [1] relies on a truncated stick-breaking prior distribution to identify the optimal number of clusters. We use the following values for the hyperparameters: $N_0 = 1000$, $\gamma_1 = \cdots = \gamma_S = 0.1$ and $\mathbf{T}$ is a diagonal matrix where the diagonal elements are equal to 10.

In our experiments, we use the Gibbs sampler implementation based on the conditional distributions described in Section 2.1 with one exception: the samples from $\Phi$ were generated using the model without covariates described in [1]. We took this approach because preliminary results revealed that this model had difficulty estimating the $\Phi$ matrix even in situations where the model without covariates estimated this matrix well. This problem arises because, differently from a standard regression in which the response variable is observed, the response variable here is latent and has to be estimated together with the regression parameters. As a result, a misspecified regression model can negatively impact the latent response variable, potentially mischaracterizing the identified clusters. Our simulated data studies reveal that this two-stage estimation results in good estimates for the parameters of interest.

In each experiment, we assessed convergence by visually evaluating trace-plots of the generated MCMC chains. Part of our algorithm was developed in R while part of the code was made in C++ using the Rcpp package [6]. An R package and a tutorial on how to use the model can be found at https://github.com/gilsonshimizu/ldacov.

# 4. Simulated experiments

First, we apply our model to two simulated data sets where all parameters are known, enabling the assessment of whether the model is estimating the parameters of interest well.

## 4.1. Simulation set 1

The first simulated dataset consists of 1000 instances, 100 categories and four clusters. Four covariates are also used, where each covariate explains only one of the four clusters, that is, the matrix with the regression slope coefficients is an identity matrix. We also choose covariate values such that some elements of the $\Theta$ are equal to 1 (i.e., some instances have elements from only one cluster). Similarly, we assume that some categories are only present in a single cluster. We do this to help model identifiability. Figures 2 (a) and (b) show a high correlation between the true and estimated elements of $\Phi$ and $\Theta$, indicating that the true parameter values can be recovered from the model when it is estimated using the strategy described in Section 3.

Table 1 shows the posterior means of the regression parameters $\boldsymbol{\beta}_k$, as well as an indicator (*) of whether their respective 99% credible intervals did not contain the value 0. In all cases, the true parameter values are contained in the corresponding credible intervals, demonstrating that our model estimates well the parameters of interest. We will omit here and in the other examples, but trace-plots of the log likelihood and model parameters (see Appendix D) demonstrated the convergence of the algorithm.

Figures 2 (c) and (d) show that the model proposed by Robert manages to estimate the $\Theta$ matrix well but has difficulty in estimating the $\Phi$ matrix.

Our method can be used to make predictions for the abundance matrix on the data samples using the information given by the covariates. Figure 3 shows that the method leads to high prediction accuracy on a hold-out set with 1000 instances.

**Table 1**
Posterior mean for the regression parameters of the simulated dataset 1.

| | Cluster 1 | | Cluster 2 | | Cluster 3 | | Cluster 4 | |
|---|---|---|---|---|---|---|---|---|
| | True | Estimated | True | Estimated | True | Estimated | True | Estimated |
| Intercept | 1.592 | 2.131* | 1.872 | 2.055* | 1.755 | 2.127* | 1.860 | 1.827* |
| Var 1 | 1.000 | 0.787* | 0.000 | −0.055 | 0.000 | −0.044 | 0.000 | −0.027 |
| Var 2 | 0.000 | 0.021 | 0.000 | 0.008 | 1.000 | 0.883* | 0.000 | 0.072 |
| Var 3 | 0.000 | −0.030 | 1.000 | 0.870* | 0.000 | −0.043 | 0.000 | −0.040 |
| Var 4 | 0.000 | −0.018 | 0.000 | 0.018 | 0.000 | 0.054 | 1.000 | 1.015* |

* "Statistically significant" results, defined as parameters for which the 99% credible intervals did not overlap zero.

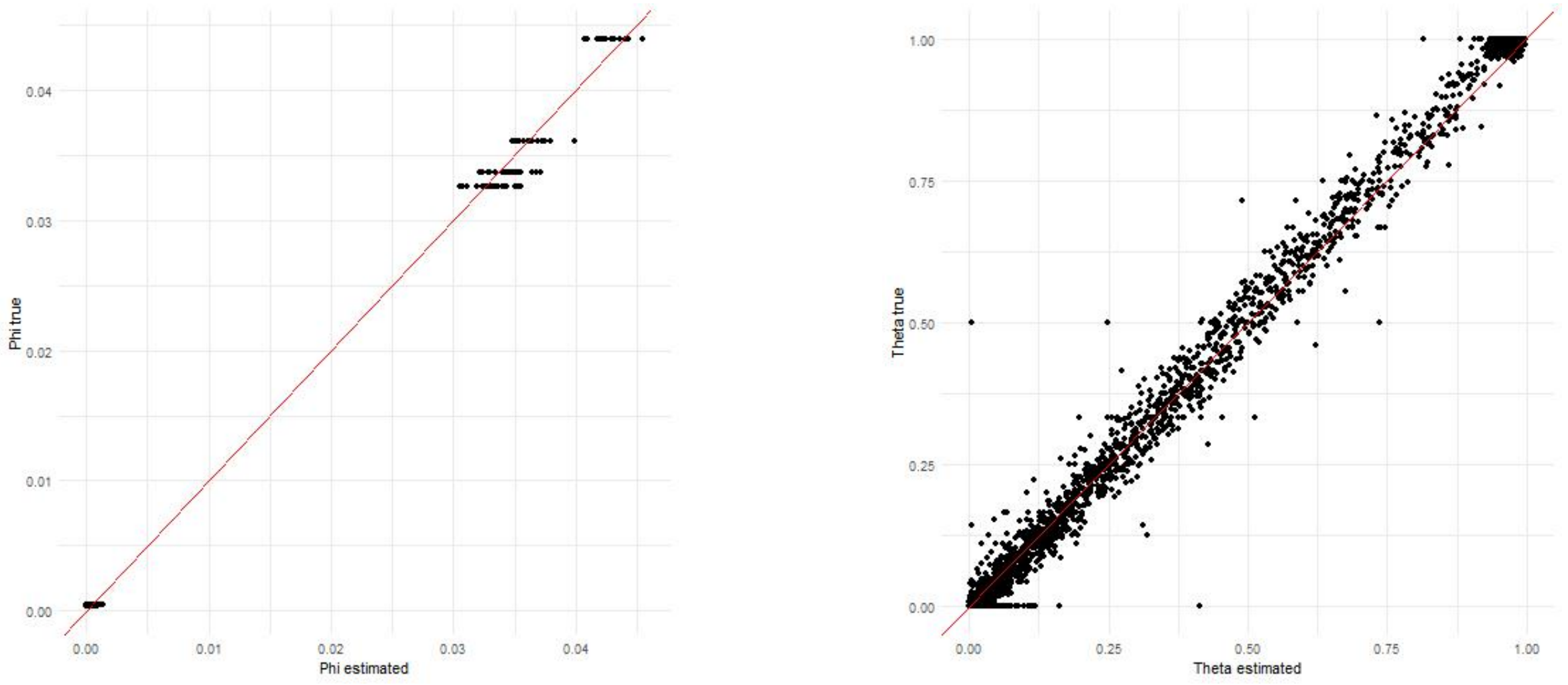

(a) Φ - new LDA (b) Θ - new LDA

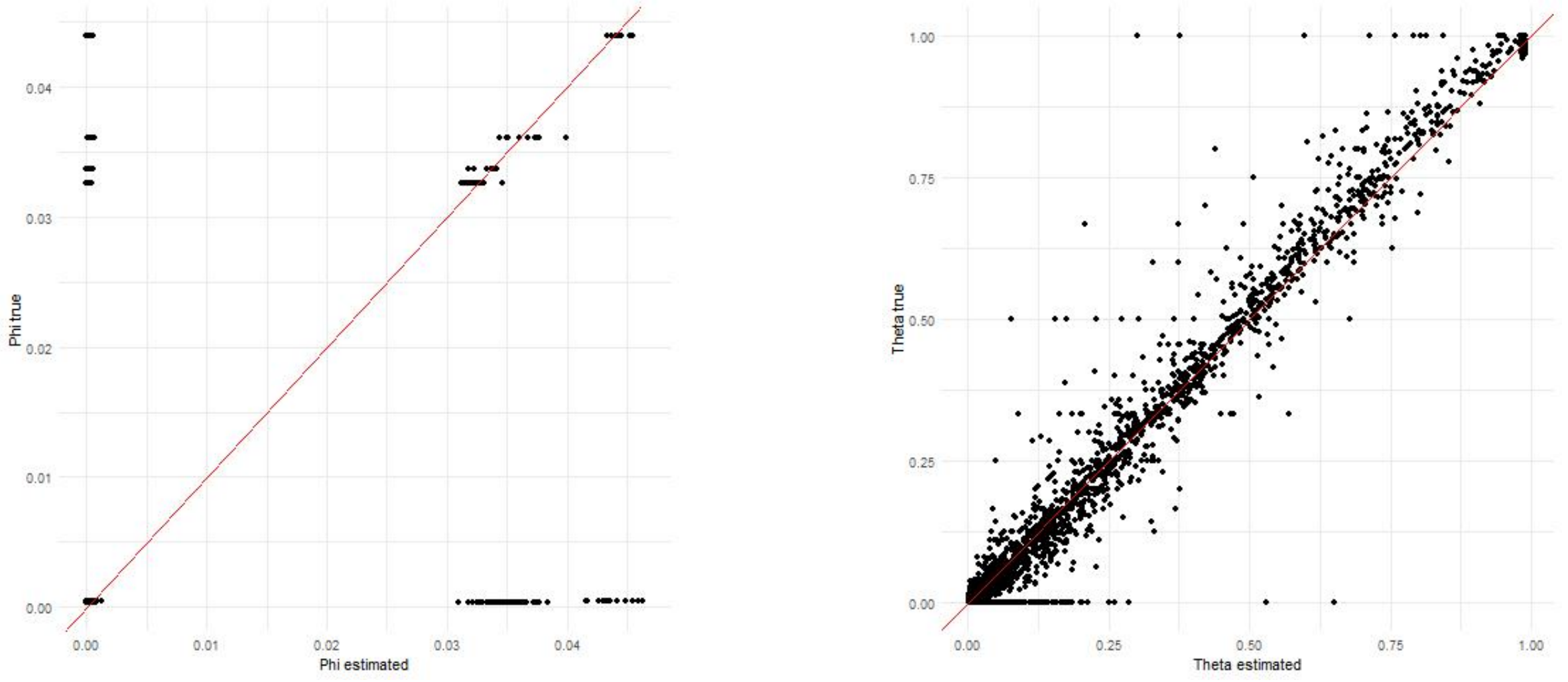

(c) Φ - STM (d) Θ - STM

**Figure 2:** Scatter plots of true and estimated values of the parameters Φ and Θ for the simulated data set 1 using new LDA formulation and STM. (a) Φ - new LDA. (b) Θ - new LDA. (c) Φ - STM. (d) Θ - STM.

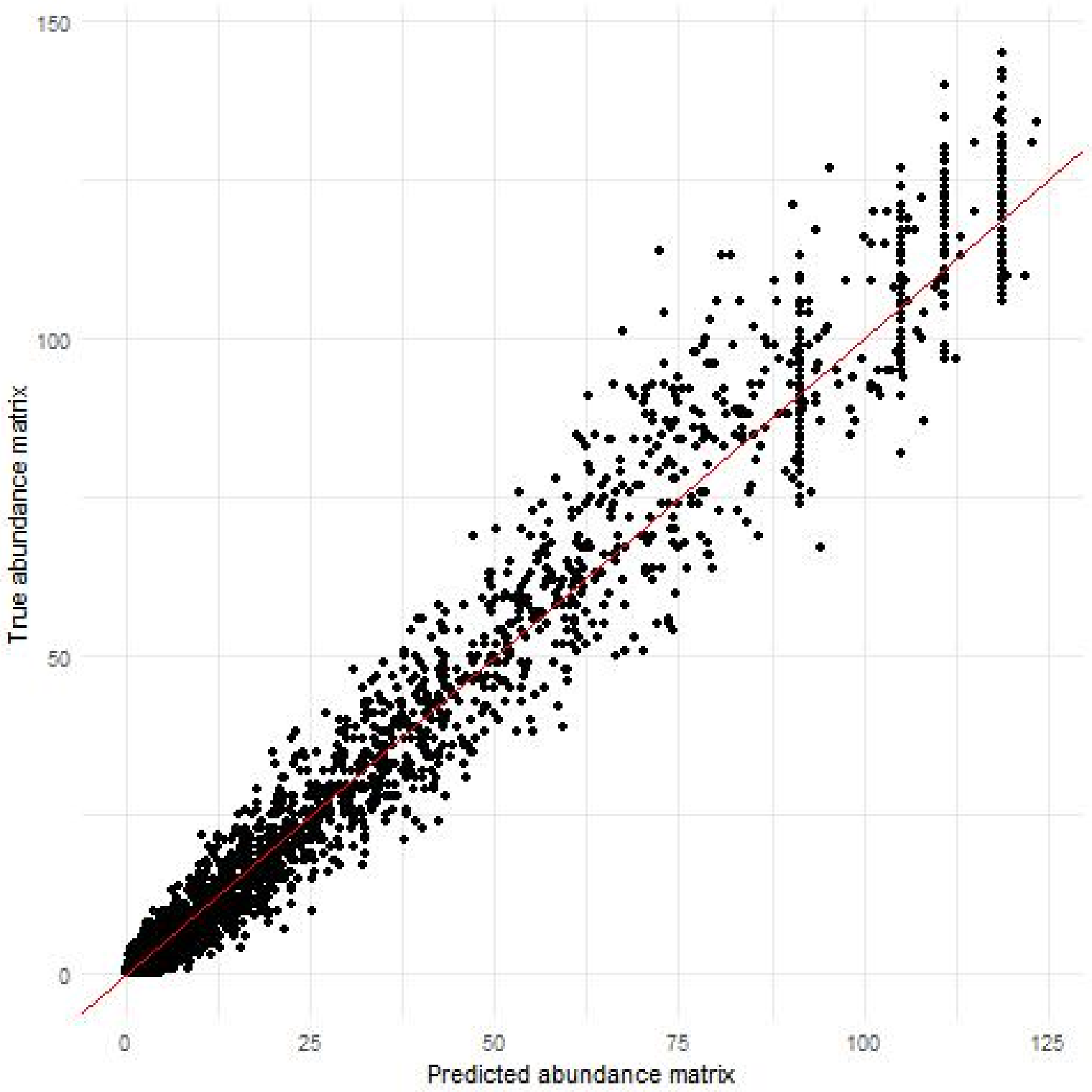

**Figure 3:** Scatter plot of the predicted abundance matrix versus true abundance matrix for the simulated data set 1.

### 4.2. Simulation set 2

The second set of simulated data is similar to the one described previously. However, instead of using the correct set of covariates, we relied on randomly generated covariates. As result, these covariates were independent of the number of individuals in each cluster. The purpose of this data set is to verify whether the model is able to infer when none of the covariates are relevant.

Figures 4 (a) and (b) show that both $\Phi$ and $\Theta$ were well estimated. Table 2 shows that all 99% credible intervals for $\beta$'s contained the value zero, which are the correct values given that the covariates were independent of the number of elements in each cluster.

Again, the model proposed by Roberts et al. [19] manages to estimate the $\Theta$ matrix well but has difficulties in estimating the $\Phi$ matrix (Figures 4 (c) and (d)).

**Table 2**
Posterior mean for the regression parameters of the simulated dataset 2.

| | Cluster 1 | | Cluster 2 | | Cluster 3 | | Cluster 4 | |
|---|---|---|---|---|---|---|---|---|
| | True | Estimated | True | Estimated | True | Estimated | True | Estimated |
| Var 1 | 0.000 | 0.046 | 0.000 | -0.052 | 0.000 | -0.009 | 0.000 | -0.002 |
| Var 2 | 0.000 | 0.018 | 0.000 | -0.059 | 0.000 | 0.028 | 0.000 | -0.018 |
| Var 3 | 0.000 | 0.110 | 0.000 | 0.082 | 0.000 | -0.102 | 0.000 | 0.012 |
| Var 4 | 0.000 | 0.047 | 0.000 | -0.028 | 0.000 | 0.020 | 0.000 | 0.022 |

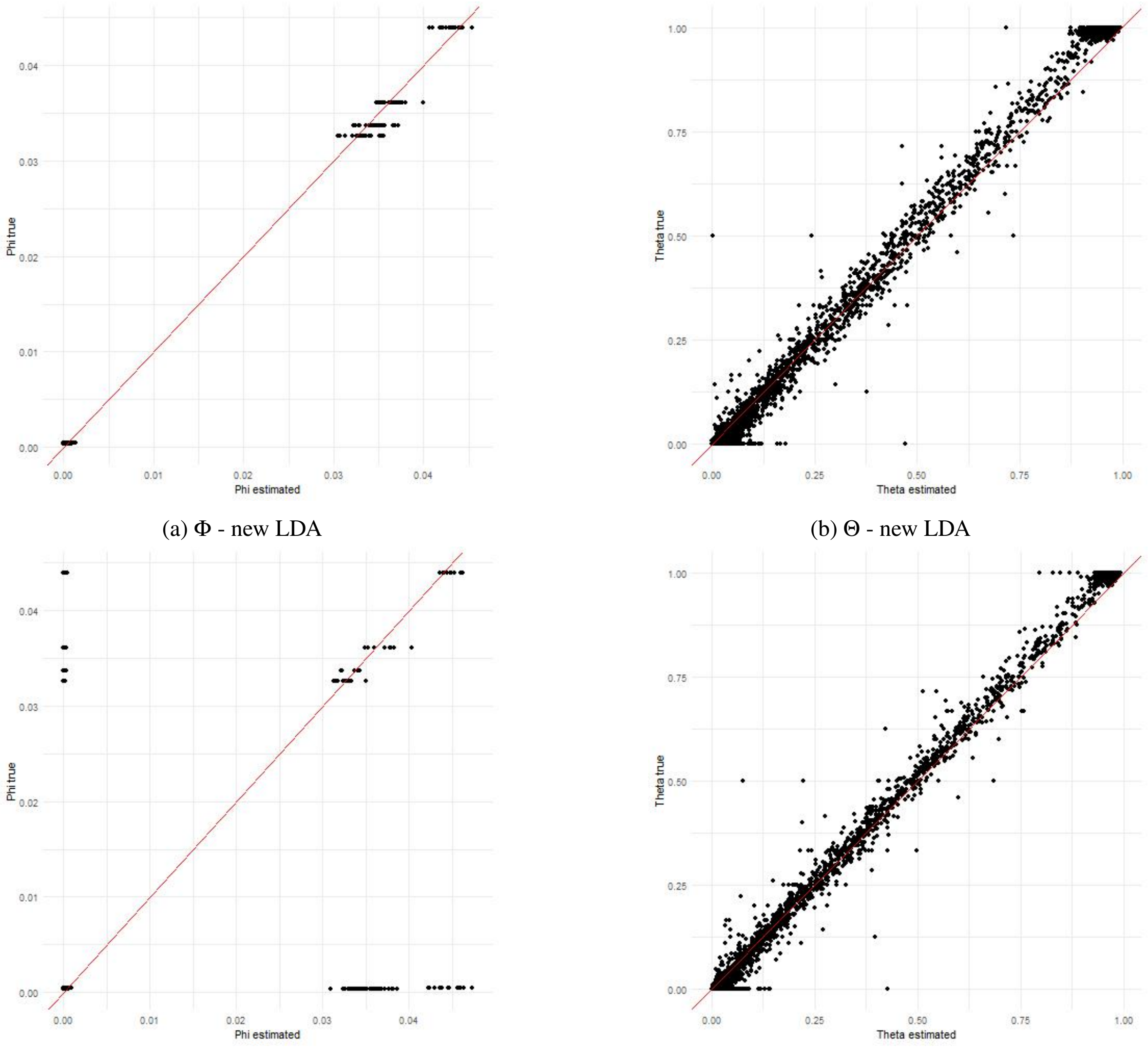

(a) $\Phi$ - new LDA (b) $\Theta$ - new LDA

(c) $\Phi$ - STM (d) $\Theta$ - STM

**Figure 4:** Test Scatter plots of true and estimated values of the parameters $\Phi$ and $\Theta$ for the simulated data set 2 using new LDA formulation and STM. (a) $\Phi$ - new LDA. (b) $\Theta$ - new LDA. (c) $\Phi$ - STM. (d) $\Theta$ - STM.

Taken together, these results reveal that our model is able to estimate the matrices $\Phi$ and $\Theta$ as well as identify the relevant variables to explain the quantities in each cluster.

## 5. Applications

To demonstrate the model's effectiveness and flexibility, we applied it to three real data sets from different areas:

- **[Covid Articles]** This dataset, available on Kaggle (`https://www.kaggle.com/allen-institute-for-ai/CORD-19-research-challenge`), contains 134,000 articles on Covid-19 and other coronavirus (on 22 april 2020). A sample of size 2,000 was extracted for analysis. We use a bag-of-words representation of the abstract in which we remove stop words, numbers and words that appear in less than 6% of abstracts. In this way, we end up with 211 words as tokens. We use the year in which the paper was published and keywords of the respective journal as covariates. The following keywords were used: virology, chemistry, infectious diseases, microbiology, veterinary, vaccine, immunology, medicine, public health and bioinformatics.

- **[Grocery Shopping]** This dataset is also available in Kaggle (`https://www.kaggle.com/karthickveerakumar/orders-data`) and contains information about 5,000 customer purchases of 99 products at a supermarket. A sample of size 2,000 was used. We use the day of the week of the purchase and the number of days since the last purchase as covariates.

- **[Barro Colorado Island]** [8] We evaluated the spatial patterns in the tree composition of the moist lowland 50-ha forest dynamic plot (FDP) on the Barro Colorado Island (BCI), Panama. FDP on BCI was established in 1981 and all free-standing woody plants with diameter at breast height (dbh) greater or equal to 1 cm were measured in 1982-93, 1985, 1990, 1995, 2000, 2005, 2010 and 2015. Annual rainfall averages 2600 mm, with a four-month dry season between December and April, while mean annual temperature is 27ºC. The total number of species identified at BCI is 326. For our analysis, we only utilized data from the last survey (2015) and we divided the FDP into 200 quadrats of size (50 m $\times$ 50 m); this was deemed the most appropriate scale to identify the spatial structure in biodiversity in BCI. We then aggregated the 2015 BCI census data by calculating the abundance of each species at each of the 200 quadrants. Before analyzing these data, we removed species that were extremely rare (defined as those species with less than 10 trees across the entire 50-ha plot). Our criteria resulted in the removal of 70 (21%) species, representing less than 0.1% of the total number of trees in our dataset.

### 5.1. Covid Articles

The number of monthly articles about coronavirus practically doubled in the first quarter of 2020 in relation to the monthly average of publications in 2019. Given this significant increase and relevance of the subject, our goal here is to find and understand the differences between possible clusters of articles. Other works involving the analysis of Twitter data and medical documents about Covid-19 can be found at [13, 16, 17]. Instead of only analyzing tabular information from articles, our analysis complements the work of Liu et al. [13] by extracting information from the abstracts of these articles.

In this text-mining application, we follow the literature in referring to topics instead of clusters. We set the maximum number of topics to 10 and use the TSB prior model proposed by Albuquerque et al. [1] to identify the optimal number of topics. This analysis identifies that the optimal number of topics was equal to 5 for this dataset.

Table 3 shows the most relevant words found in each topic. A word is defined to be relevant for a given topic if it appears at least twice as frequently in this topic than in any other topic. Table 4 shows the estimated regression weights for each topic.

Based on these results we can describe the topics found as follows:

- Topic 1: This topic is related to virology and genetics. Furthermore, this topic is more common in older articles.
- Topic 2: This topic is related to recent public health articles. This topic seems to be strongly associated with the current pandemic given that the prominent words were China, countries, outbreak, prevention, etc.
- Topic 3: This topic is related to recent articles on infectious diseases and medicine. The most relevant words in this topic suggest that it is related to tests and symptoms of Covid-19.
- Topic 4: This topic is related to older articles on vaccines, immunology and veterinary. Words like "mice" and "animal" indicate vaccine tests on non-human animals.
- Topic 5: This topic is related to older articles on viruses and influenza.

Topics 2 and 3 are more related to Covid-19 and, as expected, are strongly associated with diagnostic tests, symptoms, public health and prevention since, at the time this dataset was retrieved, there were still no vaccines or in-depth genetic studies on Covid-19. Topics 1 and 4 are related to older articles and are focused on types of studies that had not been conducted for Covid-19 at the time these data were gathered: vaccines, animal tests, and genetic studies. Topic 5 is focused on other viruses.

Although it is natural that there is a strong relationship between keywords and topics, we emphasize that the use of these keywords as covariates allowed an easier interpretation of topics than just analyzing the relevant words of each topic.

**Table 3**
Relevant words in topics of the Covid dataset.

| Topic 1 | Topic 2 | Topic 3 | Topic 4 | Topic 5 |
|---|---|---|---|---|
| protein | health | patients | mice | influenza |
| rna | public | symptoms | vaccine | viruses |
| expression | china | positive | responses | human |
| mechanism | countries | lower | levels | virus |
| sequence | outbreak | acute | evaluated | assay |
| replication | epidemic | age | groups | strains |
| target | prevention | collected | animals | highly |
| genes | emerging | samples | group | detection |
| species | diseases | confirmed | induced | |
| host | transmission | without | response | |
| genome | spread | common | immune | |
| mechanisms | research | respectively | increased | |
| molecular | future | tested | antibodies | |
| antiviral | information | performed | significantly | |
| small | population | severe | effects | |
| shown | infectious | patient | higher | |
| revealed | use | associated | caused | |
| function | strategies | hospital | type | |
| furthermore | will | detected | observed | |
| involved | current | total | compared | |
| previously | effective | clinical | effect | |
| cell | since | syndrome | significant | |
| specific | care | among | | |
| thus | number | diagnosis | | |
| | evidence | sars | | |
| | review | rate | | |
| | pathogen | | | |
| | available | | | |
| | data | | | |
| | new | | | |
| | risk | | | |

**Table 4**
Estimated Regression parameters of the covid dataset.

| | Topic 1 | Topic 2 | Topic 3 | Topic 4 | Topic 5 |
|---|---|---|---|---|---|
| Intercept | 2.876* | 2.078* | 1.510* | 0.041 | −0.256* |
| Year 2020 | −0.658* | 0.468* | 0.475* | −1.044* | −3.227* |
| Virology | 0.462* | −0.823* | −0.168 | 0.549* | 0.835* |
| Chemistry | 0.093 | −1.055* | −0.688* | −3.455* | −0.166 |
| Infect disease | −0.468* | 0.410* | 0.910* | −0.281 | 0.814* |
| Microbiology | 0.250* | −0.302* | −0.175 | −0.031 | 0.525* |
| Veterinary | −0.065 | 0.169 | 0.206 | 0.535* | −0.607* |
| Vaccine | 0.054 | −0.109 | −0.178 | 1.602* | 0.697* |
| Immunology | 0.135 | −0.643* | −0.312* | 0.825* | −0.931* |
| Medicine | −0.436* | 0.501* | 0.507* | −0.589* | 0.638* |
| Public Health | −0.969* | 1.066* | 0.137 | −0.223 | −0.722* |
| Comput bioinformatic | 0.049* | 0.586* | −3.765* | −3.037* | −0.749 |

* "Statistically significant" results, defined as parameters for which the 95% credible intervals did not overlap zero.

## 5.2. Grocery Shopping

Our goal is to find clusters of grocery shopping baskets while also identifying how these clusters are associate with day of week and days since last purchase. Other works such as [4, 9] also analyze this type of data using the LDA model.

After running the model without covariates, we obtain an ideal number of clusters equal to 4. Table 5 and Table 6 show the relevant products of each cluster and the estimates of the regression parameters, respectively.

Below we describe the clusters that were found:

- Cluster 1: This cluster has many products with purchases made on any day of the week and with varying frequencies. The products are of daily use like breads, cereals, coffee and also cleaning products.
- Cluster 2: This cluster contains herbs, spices, vegetables, poultry, etc. This type of purchase might be associated with the preparation of a special meal. This purchase is usually made on a Saturday.
- Cluster 3: This cluster contains frozen meals and prepared soups.
- Cluster 4: This cluster is very peculiar with baby formulas, beers and wines. This cluster is reflective of the classic example of basket analysis, where parents go to buy baby diapers and take the opportunity to buy beer and wine. These purchases are made less frequently than other clusters and are generally not made on Saturdays.

We note here that a customer can buy more than one basket (cluster) at the same time, which is a very interesting feature of the LDA models (compared to traditional cluster analysis) and more realistic for this data set. In a conventional cluster analysis a customer would be classified into just one cluster.

**Table 5**
Relevant products in clusters of the grocery dataset.

| Cluster 1 | Cluster 2 | Cluster 3 | Cluster 4 |
|---|---|---|---|
| cereal | fresh herbs | frozen vegan vegetarian | red wines |
| ice cream ice | canned jarred vegetables | frozen meals | baby food formula |
| water seltzer sparkling water | spices seasonings | frozen breakfast | beers coolers |
| candy chocolate | fresh vegetables | tofu meat alternatives | |
| refrigerated | poultry counter | frozen pizza | |
| frozen appetizers sides | asian foods | fresh dips tapenades | |
| tea | grains rice dried goods | energy granola bars | |
| packaged produce | canned meals beans | prepared soups salads | |
| laundry | oils vinegars | | |
| paper goods | specialty cheeses | | |
| chips pretzels | pickled goods olives | | |
| coffee | dry pasta | | |
| frozen meat seafood | packaged poultry | | |
| bread | | | |
| cleaning products | | | |
| lunch meat | | | |
| spreads | | | |
| soap | | | |
| dish detergents | | | |
| soft drinks | | | |
| ... | | | |

**Table 6**
Estimated regression parameters of the grocery dataset.

| | Cluster 1 | Cluster 2 | Cluster 3 | Cluster 4 |
|---|---|---|---|---|
| Intercept | 1.864* | 0.562* | −1.800* | −3.386* |
| Last order >= 6 days | 0.219 | 0.051 | −0.070 | 0.974* |
| Saturday | 0.152 | 0.376* | 0.330 | −0.671* |

* "Statistically significant" results, defined as parameters for which the 95% credible intervals did not overlap zero.

## 5.3. Barro Colorado Island

We ran the LDA model without covariates and, from a total of 20 potential groups, found 11 dominant groups that together comprised approximately 91% of all individuals. We find relatively strong spatial patterns in the distribution of these groups (Figure 5). For instance, group 10 is clearly restricted to the areas with steep slopes while group 3 has much higher abundance in flat areas. Interestingly, several of the groups identified here seems to closely correspond to the BCI habitat classification proposed by Harms et al. [8]. For example, group 11 seems to match the "old forest, swamp" class while group 10 seems to match the "Old forest, Streamside". However, different from this discrete classification of habitats, we find spatial patterns that reflect substantial mixed membership.

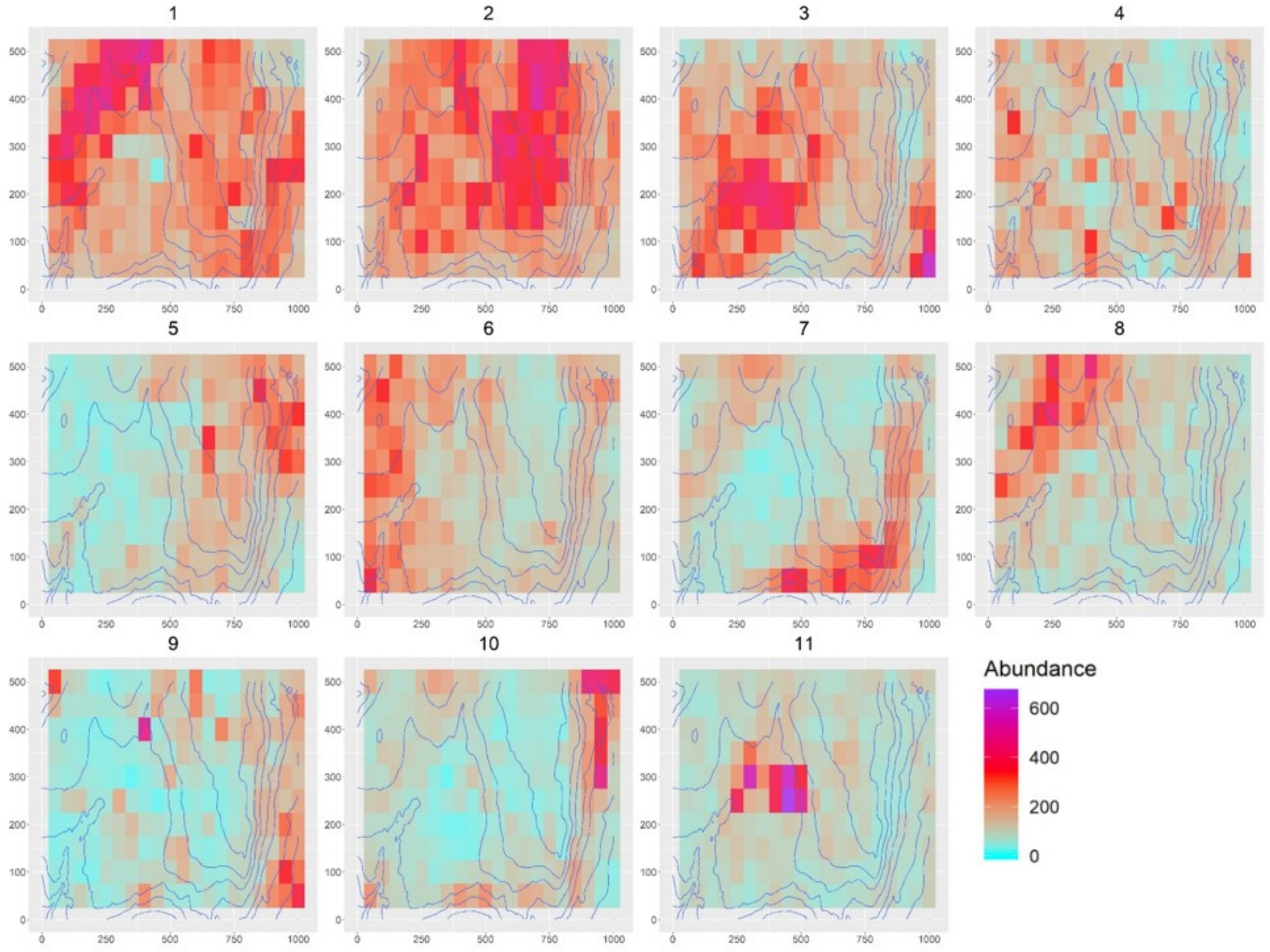

**Figure 5:** Spatial distribution of the groups identified by our model. Each panel displays the results for a given group. Hotter colors indicate higher abundance. Elevation is shown with level curves at 5-m intervals.

Similar results could have been obtained from the LDA model without covariates. The novelty of the proposed model is the ability to make formal inference on the effect of covariates (Table 7). We find that all groups, except for group 2, were strongly associated with one or more covariates. For example, as expected, groups 10 and 3 were positively and negatively associated with slope, respectively. The variables that tended to influence a large number of groups were slope, convexity, magnesium and zinc.

**Table 7**
Estimated regression parameters of the BCI dataset.

| | Groups | | | | | | | | | | |
|---|---|---|---|---|---|---|---|---|---|---|---|
| Covariates | 1 | 2 | 3 | 4 | 5 | 6 | 7 | 8 | 9 | 10 | 11 |
| Intercept | 5.474* | 5.524* | 5.238* | 4.926* | 4.678* | 4.974* | 4.746* | 4.829* | 4.529* | 4.541* | 4.645* |
| Elevation | 0.107 | 0.13 | −0.337* | −0.114 | 0.296* | 0.016 | 0.008 | 0.227* | 0.054 | 0.203* | 0.009 |
| Slope | −0.060 | −0.093 | −0.239* | 0.028 | 0.125 | 0.156* | 0.274* | 0.032 | 0.143* | 0.337* | −0.073 |
| Convexity | −0.088 | −0.03 | 0.138* | 0.094* | −0.043 | 0.086 | −0.038 | −0.129* | 0.101* | −0.153* | −0.131* |
| Al | −0.091 | 0.06 | 0.129 | −0.009 | −0.118 | −0.090 | −0.095 | −0.129 | 0.021 | −0.187* | 0.043 |
| Mn | 0.186* | −0.027 | −0.150* | 0.021 | −0.047 | 0.084 | 0.303* | 0.140* | −0.137 | 0.168* | −0.245* |
| Zn | −0.094 | −0.016 | 0.013 | −0.116 | 0.232* | −0.217* | −0.259* | −0.206* | 0.451* | −0.041 | 0.105 |
| N | 0.039 | 0.005 | 0.071 | 0.030 | −0.132 | −0.145* | 0.114 | 0.000 | −0.011 | −0.226* | −0.029 |
| pH | −0.126 | 0.080 | 0.009 | −0.034 | 0.194 | −0.099 | 0.022 | −0.277* | −0.136 | −0.002 | −0.071 |

* "Statistically significant" results, defined as parameters for which the 95% credible intervals did not overlap zero.

## 6. Model comparison using probabilistic coherence

Up to this point, we have compared STM to our model using simulated data because it enables us to determine how well these approaches estimate the true parameter values. The problem is that this approach implicitly assumes that the data follow our generative model. To avoid this assumption, we can also compare our model to STM [19] using a measure of the quality of each cluster, known as *probabilistic coherence*, as proposed by Jones [11].

The probabilistic coherence calculates for each pair of categories the measure $P(s_1|s_2) - P(s_1)$, where category $s_1$ is more likely than category $s_2$ in the focus cluster. Note that $P(s_1|s_2) - P(s_1)$ will be close to zero if $s_1$ and $s_2$ are independent and, as a result, probabilistic coherence measures how strongly associated are categories $s_1$ and $s_2$. A well delineated cluster with a high frequency of these categories should have a high probabilistic coherence. By definition, we consider only the top $M$ most frequent categories in the clusters and use the sum of the probabilistic coherence of all clusters as a measure of the quality of the estimated $\Phi$ matrix. For a specific cluster $k$, the probabilities $P(s_1|s_2)$ and $P(s_1)$ are estimated by $\frac{|\text{number of instances of categories } s_1 \text{ and } s_2 \text{ in } k|}{|\text{number of instances of category } s_2 \text{ in } k|}$ and $\frac{|\text{number of instances of category } s_1 \text{ in } k|}{|\text{number of instances in } k|}$, respectively. We used the textmineR package [10] in R to calculate probabilistic coherence with the default value of M (5).

Table 8 shows these measures for all datasets analyzed in this work. In all cases, the proposed model obtained better probabilistic coherence measures than the STM model, indicating that the topics that are found with our method are more coherent.

**Table 8**
Probabilistic coherence for all datasets comparing LDA with covariates and STM. Larger values are better and best values are highlighted in bold.

| | Dataset | | | | |
|---|---|---|---|---|---|
| Method | Simulation set 1 | Simulation set 2 | Covid Articles | Grocery Shopping | BCI |
| LDA Covariates | **1.554** | **1.550** | **0.713** | **0.370** | **0.021** |
| STM | 0.395 | 0.455 | 0.592 | 0.235 | 0.011 |

## 7. Discussion

We propose a new formulation for the LDA model that allows the incorporation of covariates. This model differs from other LDA methods because it models how covariates affects the number of elements of each cluster rather than the proportions of the clusters. Because these proportions can be derived from the number of elements, our model generalizes existing LDA models that also incorporate covariates.

The main advantage of our model is that it enables a much more straight-forward interpretation of the regression coefficients. This is due to the use of the logarithmic link function on the quantities in each cluster instead of a multinomial logistic function on the proportions. Furthermore, by more faithfully representing uncertainty, inference on the regression coefficients is likely to be better with our Gibbs sampler algorithm than when using approximate variational estimation methods.

In our simulated examples, we are able to show that our model estimates well the $\Phi$ and $\Theta$ matrices with or without relevant covariates. We also illustrate the model's ability to make inferences about regression coefficients through credible intervals and the ability to make predictions for the abundance matrix on the data samples using the information given by the covariates. Importantly, our examples with real data sets demonstrate the flexibility of the model to be applied in different areas and for different types of data.

The dataset about Covid articles, for example, is a traditional text mining data set. The use of covariates, together with the main words of each topic, enabled us to determine how the focus of these articles has changed as the new coronavirus spurs a pandemic across the world. In the data set on supermarket purchases our model was able to create clusters and relate them to the time and day of the week covariates. We believe that this tool is very useful to segment customers and also to optimize the layout of products within a grocery store. Finally, when applied to the BCI dataset, our model was able to find clusters with distinct spatial patterns and at the same time relate these patterns to some of the soil and topography features.

A disadvantage of our method is the computational cost (especially in large datasets with many sample units and categories) compared to models using variational Bayes methods. A possible future work would be to consider a version of this model with a variational inference approach.

## Acknowledgement

This study was financed in part by the Coordenação de Aperfeiçoamento de Pessoal de Nível Superior - Brasil (CAPES) - Finance Code 001.

Rafael Izbicki is grateful for the financial support of FAPESP (grant 2019/11321-9) and CNPq (grant 309607/2020-5).

This work was partly supported by the US Department of Agriculture National Institute of Food and Agriculture McIntire–Stennis project 1005163 and the US National Science Foundation award 1458034 to DV.

## A. Full Conditional Distribution of $z_{i,l}$

For simplicity, we consider only 2 clusters ($K = 2$). Suppose that our focus is on $l = l'$ and $s = s'$. It is also assumed that after removing the $i$-th element we have $[n_{l',*,1}, n_{l',*,2}]$, $[n_{l',1,1}, \dots, n_{l',S,1}]$ and $[n_{l',1,2}, \dots, n_{l',S,2}]$. We consider $\lambda_{l,k} = \exp(\mathbf{x}_l^T \boldsymbol{\beta}_k)$ and $p_{l,k} = \frac{N}{N+\lambda_{l,k}}$. We can integrate out $\boldsymbol{\phi}_k$ from

$$
\begin{aligned}
&p(z_{i,l'} = 1 \mid y_{i,l'} = s', \dots) \\
&\propto \prod_{k=1}^{K} \left[ NB(n_{l,*,k} \mid \lambda_{l,k}, N) \int \left( \prod_{l=1}^{L} Multinomial([n_{l,1,k}, \dots, n_{l,S,k}] \mid n_{l,*,k}, \boldsymbol{\phi}_k) \right) \text{Dirichlet } \left( \boldsymbol{\phi}_k \mid \gamma \right) d\boldsymbol{\phi}_k \right].
\end{aligned}
$$

The integral involving $\boldsymbol{\phi}_k$ is available in closed form (see Appendix B). Furthermore, several elements in the equation above can be eliminated because they are constants. As a result, we obtain the following expression:

$$
\begin{aligned}
&p(z_{i,l'} = 1 \mid y_{i,l'} = s', \dots) \\
&\propto \left[ \frac{\Gamma\left(n_{l',*,1} + 1 + N\right) p_{l',1}^{N} \left(1 - p_{l',1}\right)^{(n_{l',*,1}+1)}}{\Gamma(N) \left(n_{l',*,1} + 1\right)!} \times \frac{\Gamma\left(n_{l',*,2} + N\right) p_{l',2}^{N} \left(1 - p_{l',2}\right)^{(n_{l',*,2})}}{\Gamma(N) n_{l*,*,2}!} \right] \\
&\times \left( \prod_{l \neq l'} \frac{n_{l,*,1}!}{n_{l,1,1}! \dots n_{l,S,1}!} \right) \left( \frac{(n_{l',*,1} + 1)!}{n_{l',1,1}! \dots (n_{l',s,1} + 1)! \dots n_{l',S,1}!} \right) \frac{(n_{*,s',1} + 1 + \gamma_s) \prod_{s \neq s'} (n_{*,s,1} + \gamma_s)}{\Gamma(n_{*,*,1} + 1 + \sum_s \gamma_s)} \\
&\times \frac{n_{l,*,2}!}{n_{l,1,2}! \dots n_{l,S,2}!} \frac{\prod_{s=1} (n_{*,s,2} + \gamma_s)}{\Gamma(n_{*,*,2} + \sum_s \gamma_s)}.
\end{aligned}
$$

We drop additional terms that are constants to obtain:

$$
\begin{aligned}
&p(z_{i,l'} = 1 \mid y_{i,l'} = s', \dots) \\
&\propto \left[ \frac{\Gamma\left(n_{l',*,1} + 1 + N\right)\left(1 - p_{l',1}\right)^{\left(n_{l',*,1}+1\right)}}{\left(n_{l',*,1} + 1\right)!} \times \frac{\Gamma\left(n_{l',*,2} + N\right)\left(1 - p_{l',2}\right)^{\left(n_{l',*,2}\right)}}{n_{l^*,*,2}!} \right] \\
&\times \left( \frac{(n_{l',*,1} + 1)!}{n_{l',1,1}! \dots (n_{l',s,1} + 1)! \dots n_{l',S,1}!} \right) \frac{(n_{*,s',1} + 1 + \gamma_s)}{\Gamma(n_{*,*,1} + 1 + \sum_s \gamma_s)} \\
&\times \left( \frac{n_{l',*,2}!}{n_{l',1,2}! \dots n_{l',s',2}! \dots n_{l',S,2}!} \right) \frac{\left(n_{*,s',2} + \gamma_{s'}\right)}{\Gamma\left(n_{*,*,2} + \sum_s \gamma_s\right)} \propto b_1 a_1
\end{aligned}
$$

where:

$$
a_1 = \left( \frac{(n_{l',*,1} + 1)!}{n_{l',1,1}! \dots (n_{l',s,1} + 1)! \dots n_{l',S,1}!} \right) \frac{(n_{*,s',1} + 1 + \gamma_s)}{\Gamma(n_{*,*,1} + 1 + \sum_s \gamma_s)} \times \left( \frac{n_{l',*,2}!}{n_{l',1,2}! \dots n_{l',s',2}! \dots n_{l',S,2}!} \right) \frac{\left(n_{*,s',2} + \gamma_{s'}\right)}{\Gamma\left(n_{*,*,2} + \sum_s \gamma_s\right)} \text{ and}
$$

$$
b_1 = \left[ \frac{\Gamma\left(n_{l',*,1} + 1 + N\right)\left(1 - p_{l',1}\right)^{\left(n_{l',*,1}+1\right)}}{\left(n_{l',*,1} + 1\right)!} \times \frac{\Gamma\left(n_{l',*,2} + N\right)\left(1 - p_{l',2}\right)^{\left(n_{l',*,2}\right)}}{n_{l^*,*,2}!} \right].
$$

Similarly, it can be shown that

$$
p(z_{i,l'} = 2 \mid y_{i,l'} = s', \dots) \propto b_2 a_2
$$

where

$$
a_2 = \left( \frac{n_{l',*,1}!}{n_{l',1,1}! \dots n_{l',s',1}! \dots n_{l',S,1}!} \right) \frac{\left(n_{*,s',1} + \gamma_{s'}\right)}{\Gamma\left(n_{*,*,1} + \sum_s \gamma_s\right)} \times \left( \frac{(n_{l',*,2} + 1)!}{n_{l',1,2}! \dots (n_{l',s,2} + 1)! \dots n_{l',S,2}!} \right) \frac{(n_{*,s',2} + 1 + \gamma_s)}{\Gamma(n_{*,*,2} + 1 + \sum_s \gamma_s)} \text{ and}
$$

$$
b_2 = \left[ \frac{\Gamma\left(n_{l',*,1} + N\right)\left(1 - p_{l',1}\right)^{\left(n_{l',*,1}\right)}}{n_{l^*,*,1}!} \times \frac{\Gamma\left(n_{l',*,2} + 1 + N\right)\left(1 - p_{l',2}\right)^{\left(n_{l'*,2}+1\right)}}{\left(n_{l',*,2} + 1\right)!} \right].
$$

Because $z_{i,l'}$ is either equal to 1 or 2, we can divide both by $b_1 a_1 + b_2 a_2$ and, using factorial and gamma function rules, we obtain:

$$
\begin{aligned}
p(z_{i,l'} = 1 \mid y_{i,l'} = s', \dots) &\propto \left( \frac{\frac{(n_{l',*,1}+1)(n_{*,s',1}+\gamma_{s'})}{(n_{l',s'}+1)(n_{*,*,1}+\sum_s \gamma_s)}}{\frac{(n_{l',*,1}+1)(n_{*,s',1}+\gamma_{s'})}{(n_{l',s',1}+1)(n_{*,*,1}+\sum_s \gamma_s)} + \frac{(n_{l',*,2}+1)(n_{*,s',2}+\gamma_{s'})}{(n_{l',s',2+1})(n_{*,*,2}+\sum_s \gamma_s)}} \right) \left( \frac{\frac{(n_{l',*,1}+N)(1-p_{l',1})}{(n_{l',*,1}+1)}}{\frac{(n_{l',*,1}+N)(1-p_{l',1})}{(n_{l',*,1}+1)} + \frac{(n_{l',*,2}+N)(1-p_{l',2})}{(n_{l',*,2}+1)}} \right) \\
&\propto \frac{\left(n_{l',*,1} + N\right)\left(n_{*s',1} + \gamma_{s'}\right)}{\left(n_{l's',1} + 1\right)\left(n_{*,*,1} + \sum_s \gamma_s\right)} \left(1 - p_{l',1}\right).
\end{aligned}
$$

And finally we have that

$$
z_{i,l} \mid y_{i,l} = s, \cdots \sim Categorical\left( \left[ \frac{\frac{(n_{l,*,1}+N)(n_{*,s,1}+\gamma_s)}{(n_{l,s,1}+1)(n_{*,*,1}+\sum_s \gamma_s)}(1-p_{l,1})}{\sum_{k=1}^{K} \frac{(n_{l,*,k}+N)(n_{*,s,k}+\gamma_s)}{(n_{l,s,k}+1)(n_{*,*,k}+\sum_s \gamma_s)}(1-p_{l,k})}, \dots, \frac{\frac{(n_{l,*,K}+N)(n_{*,s,K}+\gamma_s)}{(n_{l,s,K}+1)(n_{*,*,K}+\sum_s \gamma_s)}(1-p_{l,K})}{\sum_{k=1}^{K} \frac{(n_{l,*,k}+N)(n_{*,s,k}+\gamma_s)}{(n_{l,s,k}+1)(n_{*,*,k}+\sum_s \gamma_s)}(1-p_{l,k})} \right] \right).
$$

## B. Multinomial Integration in $\boldsymbol{\phi}_k$

To simplify the calculation of the conditional distribution of $z_{i,l}$ we can integrate out $\boldsymbol{\phi}_k$ as shown below.

$$
\begin{aligned}
&\prod_{k=1}^{K} \int \left[ \prod_{l=1}^{L} \text{Multinomial}\left( \left[ n_{l,1,k}, \dots, n_{l,S,k} \right] \mid n_{l,*,k}, \boldsymbol{\phi}_k \right) \right] \text{ Dirichlet } \left( \boldsymbol{\phi}_k \mid \gamma \right) d\boldsymbol{\phi}_k \\
&\propto \prod_{k=1}^{K} \int \left[ \prod_{l=1}^{L} \frac{n_{l,*,k}!}{n_{l,1,k}! \dots n_{l,S,k}!} \phi_{k,1}^{n_{l,1,k}} \dots \phi_{k,S}^{n_{l,S,k}} \right] \phi_{k,1}^{\gamma_1 - 1} \dots \phi_{k,S}^{\gamma_S - 1} d\boldsymbol{\phi}_k \\
&\propto \prod_{k=1}^{K} \left( \prod_{l=1}^{L} \frac{n_{l,*,k}!}{n_{l,1,k}! \dots n_{l,S,k}!} \right) \int \phi_{k,1}^{n_{*,1,k}+\gamma_1-1} \dots \phi_{k,S}^{n_{*,S,k}+\gamma_S-1} d\boldsymbol{\phi}_k \\
&\propto \prod_{k=1}^{K} \left( \prod_{l=1}^{L} \frac{n_{l,*,k}!}{n_{l,1,k}! \dots n_{l,S,k}!} \right) \frac{\prod_{s=1}^{S} \left(n_{*,s,k} + \gamma_s\right)}{\Gamma\left(n_{*,*,k} + \sum_{s=1}^{S} \gamma_s\right)}
\end{aligned}
$$

## C. Slice Sampling

**Algorithm 1** Slice Sampling [5]

1: Choose an initial value $x_0$ for which $f(x_0) > 0$.
2: Sample a value of $y$ uniformly between 0 and $f(x_0)$.
3: Draw a horizontal line through the curve at this $y$ position.
4: Sample a point $(x, y)$ from the line inside the curve.
5: Repeat from step 2 using the new value of $x$.

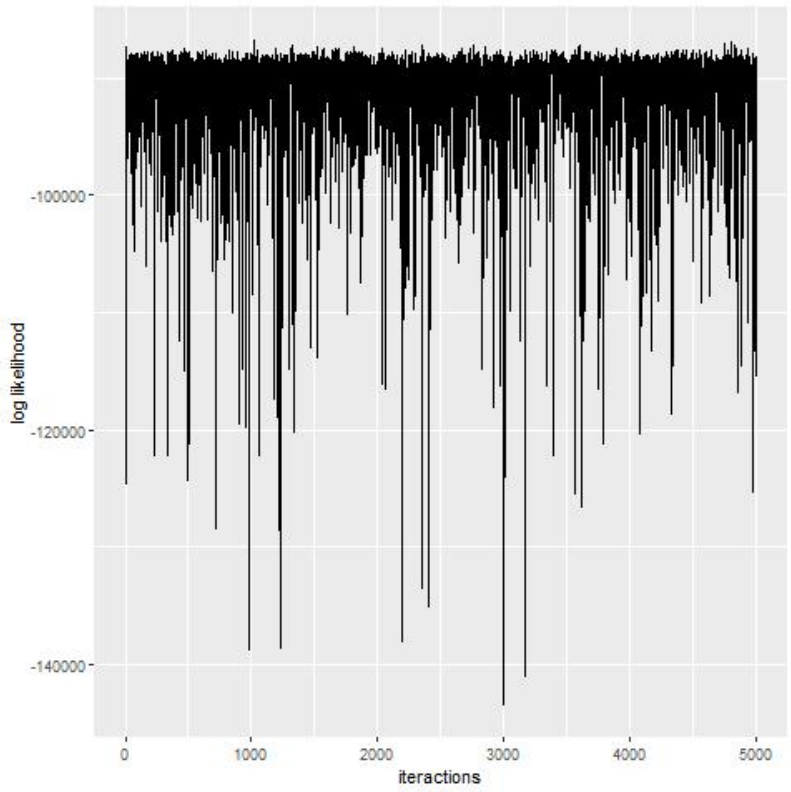

(a) Simulation 1

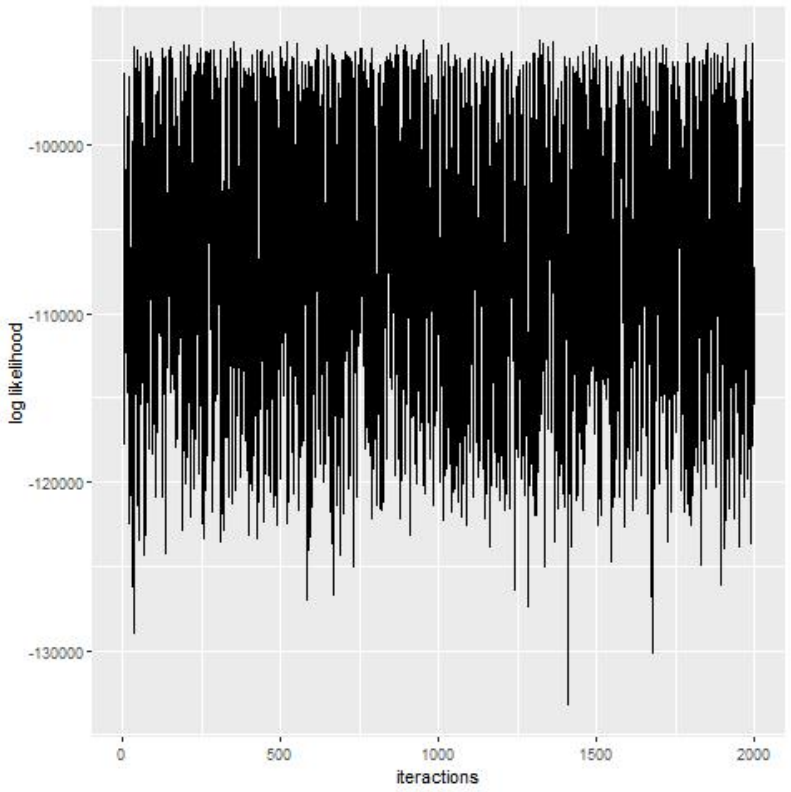

(b) Simulation 2

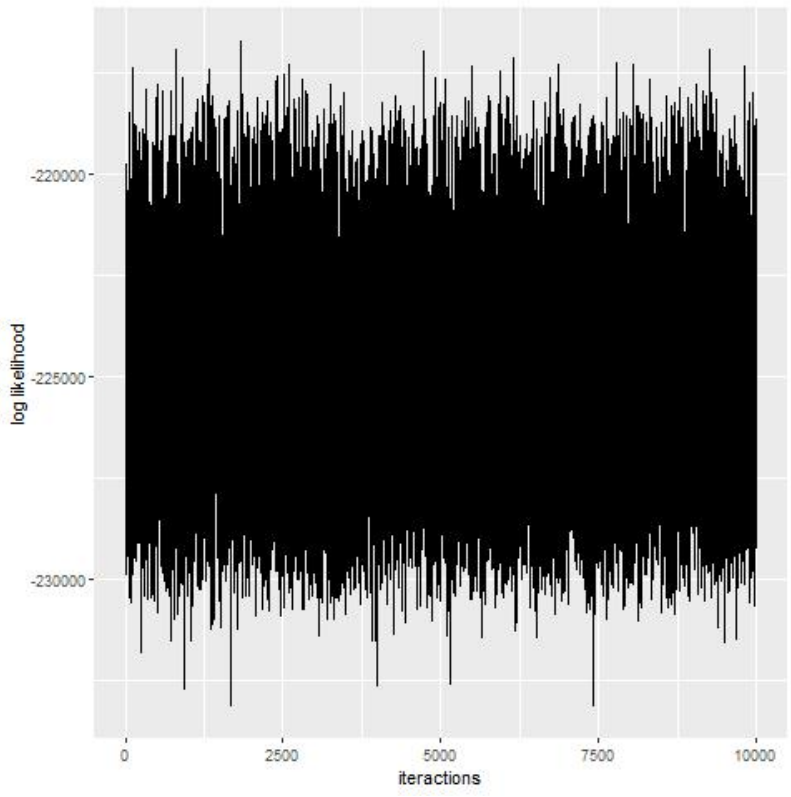

(c) Covid

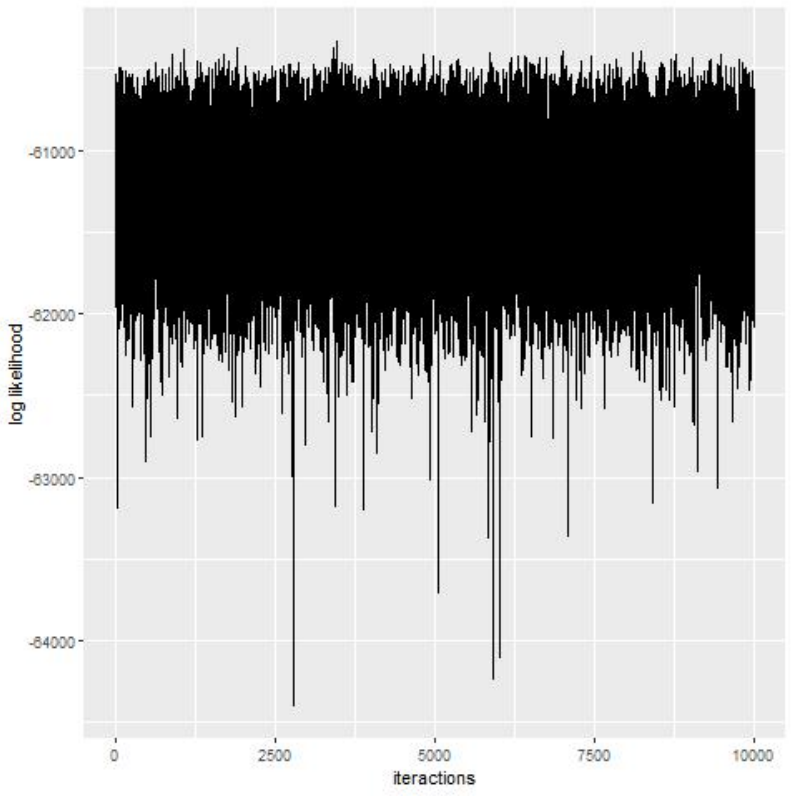

(d) Grocery

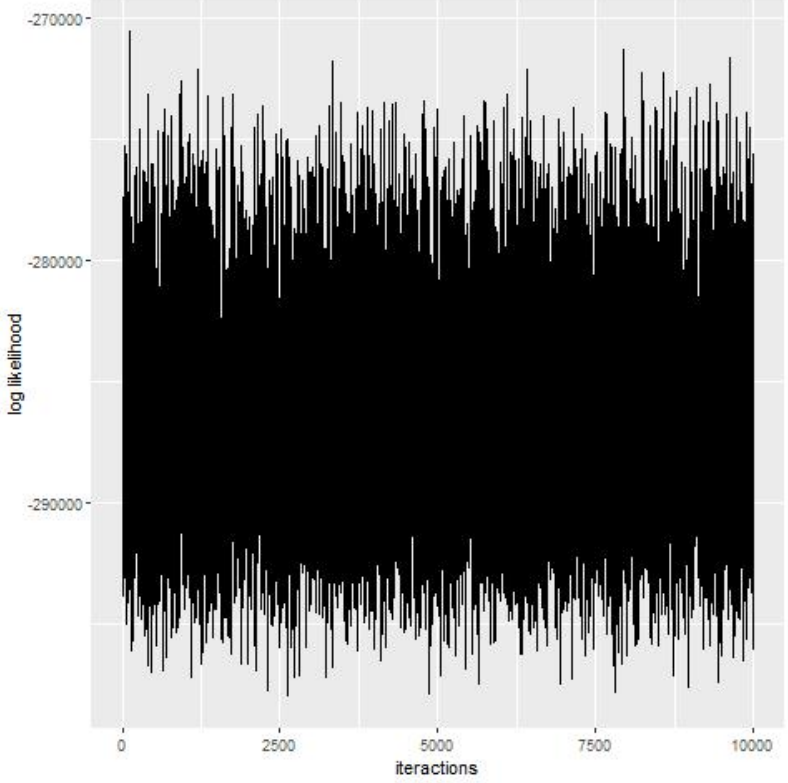

(e) BCI

**Figure 6:** MCMC convergence diagnostics of simulated and real data. (a) Simulation 1. (b) Simulation 2. (c) Covid. (d) Grocery. (e) BCI.

# D. MCMC Convergence Diagnostics

Figures 6 (a), (b), (c), (d) and (e) show the convergence diagnosis of the maximum likelihood function for all analyzed data sets.

## References

[1] Albuquerque, P.H., do Valle, D.R., Li, D., 2019. Bayesian lda for mixed-membership clustering analysis: The rlda package. Knowledge-Based Systems 163, 988–995.

[2] Blei, D.M., Lafferty, J.D., 2006. Dynamic topic models, in: Proceedings of the 23rd international conference on Machine learning, pp. 113–120.

[3] Blei, D.M., Ng, A.Y., Jordan, M.I., 2003. Latent dirichlet allocation. Journal of machine Learning research 3, 993–1022.

[4] Christidis, K., Apostolou, D., Mentzas, G., 2010. Exploring customer preferences with probabilistic topics models, in: European Conference on Machine Learning and Principles and Practice of Knowledge Discovery in Databases, pp. 12–24.

[5] Damlen, P., Wakefield, J., Walker, S., 1999. Gibbs sampling for bayesian non-conjugate and hierarchical models by using auxiliary variables. Journal of the Royal Statistical Society: Series B (Statistical Methodology) 61, 331–344.

[6] Eddelbuettel, D., François, R., Allaire, J., Ushey, K., Kou, Q., Russel, N., Chambers, J., Bates, D., 2011. Rcpp: Seamless r and c++ integration. Journal of Statistical Software 40, 1–18.

[7] Geman, S., Geman, D., 1984. Stochastic relaxation, gibbs distributions, and the bayesian restoration of images. IEEE Transactions on pattern analysis and machine intelligence , 721–741.

[8] Harms, K.E., Condit, R., Hubbell, S.P., Foster, R.B., 2001. Habitat associations of trees and shrubs in a 50-ha neotropical forest plot. Journal of Ecology 89, 947–959.

[9] Jacobs, B.J., Donkers, B., Fok, D., 2016. Model-based purchase predictions for large assortments. Marketing Science 35, 389–404.

[10] Jones, T., 2019. textmineR: Functions for Text Mining and Topic Modeling. URL: https://CRAN.R-project.org/package=textmineR. r package version 3.0.4.

[11] Jones, T., 2021. Topic modeling. https://cran.r-project.org/web/packages/textmineR/vignettes/c_topic_modeling.html. Accessed: 2021-06-27.

[12] Lienou, M., Maitre, H., Datcu, M., 2009. Semantic annotation of satellite images using latent dirichlet allocation. IEEE Geoscience and Remote Sensing Letters 7, 28–32.

[13] Liu, N., Chee, M.L., Niu, C., Pek, P.P., Siddiqui, F.J., Ansah, J.P., Matchar, D.B., Lam, S.S.W., Abdullah, H.R., Chan, A., et al., 2020. Coronavirus disease 2019 (covid-19): an evidence map of medical literature. BMC medical research methodology 20, 1–11.

[14] Lukins, S.K., Kraft, N.A., Etzkorn, L.H., 2010. Bug localization using latent dirichlet allocation. Information and Software Technology 52, 972–990.

[15] Mcauliffe, J.D., Blei, D.M., 2008. Supervised topic models, in: Advances in neural information processing systems, pp. 121–128.

[16] Ordun, C., Purushotham, S., Raff, E., 2020. Exploratory analysis of covid-19 tweets using topic modeling, umap, and digraphs. arXiv preprint arXiv:2005.03082 .

[17] Otmakhova, Y., Verspoor, K., Baldwin, T., Suster, S., 2020. Improved topic representations of medical documents to assist covid-19 literature exploration. in Proceedings of the 1st Workshop on NLP for COVID19 (Part 2) at EMNLP 2020. Online: Association for Computational Linguistics, Dec. 2020 .

[18] Pritchard, J.K., Stephens, M., Donnelly, P., 2000. Inference of population structure using multilocus genotype data. Genetics 155, 945–959.

[19] Roberts, M.E., Stewart, B.M., Airoldi, E.M., 2016. A model of text for experimentation in the social sciences. Journal of the American Statistical Association 111, 988–1003.

[20] Valle, D., Baiser, B., Woodall, C.W., Chazdon, R., 2014. Decomposing biodiversity data using the latent dirichlet allocation model, a probabilistic multivariate statistical method. Ecology letters 17, 1591–1601.

[21] Wang, X., Grimson, E., 2008. Spatial latent dirichlet allocation, in: Advances in neural information processing systems, pp. 1577–1584.

[22] Xing, D., Girolami, M., 2007. Employing latent dirichlet allocation for fraud detection in telecommunications. Pattern Recognition Letters 28, 1727–1734.

Gilson Shimizu has been a postdoctoral researcher at the Institute of Mathematics and Computer Sciences (USP, Brazil). He received his Ph.D in Statistics from Federal University of Sao Carlos (UFSCar, Brazil) with emphasis on Machine Learning methods. He has a MS in statistics (UFSCar, Brazil), an MBA from Fundaçao Getulio Vargas (FGV, Brazil) and his B.A. is from the Sao Paulo State University (UNESP, Brazil) in statistics. His experience includes 13 years as specialist in machine learning methods applied to the marketing, insurance and financial industry.

Since 2014, Rafael Izbicki has been a faculty at the Department of Statistics at the Federal University of Sao Carlos (UFSCar, Brazil). He received his Ph.D in Statistics from Carnegie Mellon University (CMU, USA) and his B.A. from University of Sao Paulo (IME-USP, Brazil). He has published several papers in the areas of Machine Learning, Nonparametric Statistics, Foundations of Statistics, Decision Theory, Bayesian Statistics and High-Dimensional Inference. He is one of the founders of the Statistical Machine Learning Lab at UFSCar (SMaLL/UFSCar).

Denis Valle has been a faculty at the School of Forest Resources and Conservation at the University of Florida (UF, USA) since 2013. He received his Ph.D in Ecology from Duke University (USA) and he has an MS in statistics (Duke University) and an MS in forestry (UF, USA). His B.A is from the University of Sao Paulo (ESALQ-USP, Brazil). He has published several papers on novel Bayesian methods applied to Ecology and conservation.